\documentclass[acmsmall]{acmart}
\usepackage{xspace}

\newcommand{\modell}{\textsf{RecRanker}}

\newcommand{\model}{\textsf{RecRanker}\xspace}

\usepackage{dsfont}
\usepackage{multirow}
\usepackage{enumitem}
\usepackage[ruled,vlined,linesnumbered,lined, commentsnumbered]{algorithm2e}
\usepackage[noend]{algpseudocode}

\usepackage{bm}
\usepackage{subfig}
\usepackage{adjustbox}
\usepackage{hyperref}

\AtBeginDocument{%
  \providecommand\BibTeX{{%
    \normalfont B\kern-0.5em{\scshape i\kern-0.25em b}\kern-0.8em\TeX}}}

\author{Sichun Luo}
\affiliation{
  \institution{City University of Hong Kong}
  \city{Hong Kong}
  \state{Kowloon}
  \country{China}
}
\email{sichun.luo@my.cityu.edu.hk}

\author{Bowei	He}
\affiliation{
  \institution{City University of Hong Kong}
  \city{Hong Kong}
  \state{Kowloon}
  \country{China}
}
\email{boweihe2-c@my.cityu.edu.hk}

\author{Haohan	Zhao}
\affiliation{
  \institution{Chinese Academy of Sciences, Hong Kong Institute of Science \& Innovation}
  \city{Hong Kong}
  \state{Kowloon}
  \country{China}
}
\email{haohazhao2-c@my.cityu.edu.hk}

\author{Wei	Shao}
\affiliation{
  \institution{City University of Hong Kong}
  \city{Hong Kong}
  \state{Kowloon}
  \country{China}
}
\email{weishao4-c@my.cityu.edu.hk}

\author{Yanlin	Qi}
\affiliation{
  \institution{Harbin Institute of Technology, Shenzhen}
  \city{Shenzhen}
  \state{Guangdong}
  \country{China}
}
\email{yanlinqi7@gmail.com}

\author{Yinya	Huang}
\affiliation{
  \institution{City University of Hong Kong}
  \city{Hong Kong}
  \state{Kowloon}
  \country{China}
}
\email{yinya.el.huang@gmail.com}

\author{Aojun	Zhou}
\affiliation{
  \institution{Chinese University of Hong Kong}
  \city{Hong Kong}
  \state{Kowloon}
  \country{China}
}
\email{aojunzhou@gmail.com}

\author{Yuxuan	Yao}
\affiliation{
  \institution{City University of Hong Kong}
  \city{Hong Kong}
  \state{Kowloon}
  \country{China}
}
\email{yuxuanyao3-c@my.cityu.edu.hk}

\author{Zongpeng	Li}
\affiliation{
  \institution{Hangdian University}
  \city{Hangzhou}
  \state{Zhejiang}
  \country{China}
  }
\email{zongpeng@tsinghua.edu.cn}

\author{Yuanzhang Xiao}
\affiliation{%
\department{Hawaii Advanced Wireless Technologies Institute}
  \institution{University of Hawaii at Manoa}
  \city{Honolulu}
  \state{HI}
  \country{USA}
}
\email{yxiao8@hawaii.edu}

\author{Mingjie Zhan}
\affiliation{
  \institution{ Sensetime Research }
  \city{ Beijing }
  \state{ Beijing }
  \country{China}
}
\email{zhanmingjie@sensetime.com}

\author{Linqi Song}
\authornote{Corresponding author.}
\affiliation{
  \institution{City University of Hong Kong}
  \city{Hong Kong}
  \state{Kowloon}
  \country{China}
  \postcode{999077}
}
\email{linqisong@cityu.edu.hk}

\renewcommand{\shortauthors}{Luo et al.}

\setcopyright{acmcopyright}

\begin{document}

\title{RecRanker: Instruction Tuning Large Language Model as Ranker for Top-k Recommendation}



\begin{abstract}

Large Language Models (LLMs) have demonstrated remarkable capabilities and have been extensively deployed across various domains, including recommender systems. Prior research has employed specialized \textit{prompts} to leverage the in-context learning capabilities of LLMs for recommendation purposes. More recent studies have utilized instruction tuning techniques to align LLMs with human preferences, promising more effective recommendations.
However, existing methods suffer from several limitations. The full potential of LLMs is not fully elicited due to low-quality tuning data and the overlooked integration of conventional recommender signals. Furthermore, LLMs may generate inconsistent responses for different ranking tasks in the recommendation, potentially leading to unreliable results. 




In this paper, we introduce \textbf{\model}, tailored for instruction tuning LLMs to serve as the \textbf{Ranker} for top-\textit{k} \textbf{Rec}ommendations. Specifically, we introduce importance-aware sampling, clustering-based sampling, and penalty for repetitive sampling for sampling high-quality, representative, and diverse training data. To enhance the prompt, we introduce a position shifting strategy to mitigate position bias and augment the prompt with auxiliary information from conventional recommendation models, thereby enriching the contextual understanding of the LLM. Subsequently, we utilize the sampled data to assemble an instruction-tuning dataset with the augmented prompts comprising three distinct ranking tasks: pointwise, pairwise, and listwise rankings. We further propose a hybrid ranking method to enhance the model performance by ensembling these ranking tasks. Our empirical evaluations demonstrate the effectiveness of our proposed \model in both direct and sequential recommendation scenarios.\footnote{Link to this paper’s source code is available at: \url{https://github.com/sichunluo/RecRanker}.}


\end{abstract}




\maketitle

\section{Introduction}
Recommender systems serve as information filtering techniques designed to mitigate the problem of information overload \cite{bobadilla2013recommender,zhang2019deep,gao2023survey}. Among various scenarios within recommender systems, the top-\textit{k} recommendation paradigm is particularly noteworthy by providing users with a list of top-$k$ items most relevant to their preferences \cite{yang2012top,zhu2019improving}. Top-\textit{k} recommendations encompass diverse tasks, including but not limited to, collaborative filtering-based direct recommendations and sequential recommendations. 
{On the one hand, direct recommendations  are studied by some prominent methodologies including}
NCF \cite{he2017neural}, NGCF \cite{wang2019neural}, and LightGCN \cite{he2020lightgcn}. These techniques harness collaborative information via neural networks.
On the other hand, for sequential recommendations \cite{chen2023sim2rec}, representative methods like SASRec \cite{kang2018self} and BERT4Rec \cite{sun2019bert4rec} utilize the attention mechanism \cite{vaswani2017attention} to model user sequences.

In recent years, Large Language Models (LLMs) \cite{openai2023gpt4,touvron2023llama2,anil2023palm} have exhibited significant prowess in natural language understanding \cite{du2022shortcut}, generation \cite{todd2023level}, and complex reasoning \cite{kojima2022large}. Consequently, they have been increasingly integrated into a multitude of domains, including recommender systems \cite{wu2023survey,fan2023recommender,lin2023can}. A typical example of LLMs in this context is to function as a ranker for a pre-filtered set of recommendations \cite{liu2023chatgpt}. 
This preference for LLMs as rankers arises primarily from the inherent limitations of LLMs, including their constrained context size and the potential for high computational costs when processing vast pools of candidate items. 
Therefore, a retrieval model is often employed to narrow down the candidate set, upon which the LLM utilizes its contextual understanding and reasoning capabilities to generate a ranked list of recommendations \cite{wang2023zero}. For example, Hou et al. \cite{hou2023large} operate LLM as a zero-shot ranker for sequential recommendation by formalizing the recommendation as a conditional ranking task based on sequential interaction histories.
By employing carefully designed prompting templates and conducting experiments on standard datasets, they show LLMs exhibit promising zero-shot ranking capabilities that can outperform traditional models. 
However, these methods possess certain limitations. The standard, general-purpose LLM does not inherently align with recommendation objectives. 
To remedy this, Zhang et al. \cite{zhang2023recommendation} suggest employing instruction tuning to better align the LLM with specific recommendation tasks. 
They express user preferences as natural language instructions, tuning the LLM to deliver more precise and user-centric recommendations. 

Nonetheless, existing methods have several limitations. 
\textit{Firstly}, current methods tend to utilize randomly sampled data of somewhat low quality for tuning LLMs, which may not fully elicit the potential of these models. The importance of high-quality training data sampling is often neglected, potentially resulting in sub-optimal model performance.
\textit{Secondly}, prevalent text-based approaches predominantly rely on the textual information of users and items for LLM processing and reasoning. The absence of integration with signals from conventional recommendation models may restrict the effectiveness of these approaches.
\textit{Finally}, LLMs may generate inconsistent responses for different ranking tasks within the recommendation process, which could lead to unreliable results. Existing research has not thoroughly explored the ranking task. Most studies deploy LLMs for a single ranking task, neglecting the potential benefits of integrating multiple ranking tasks for more reliable and trustworthy results. 





\begin{figure*}[t!]
    \centering
    \includegraphics[width=1\linewidth]{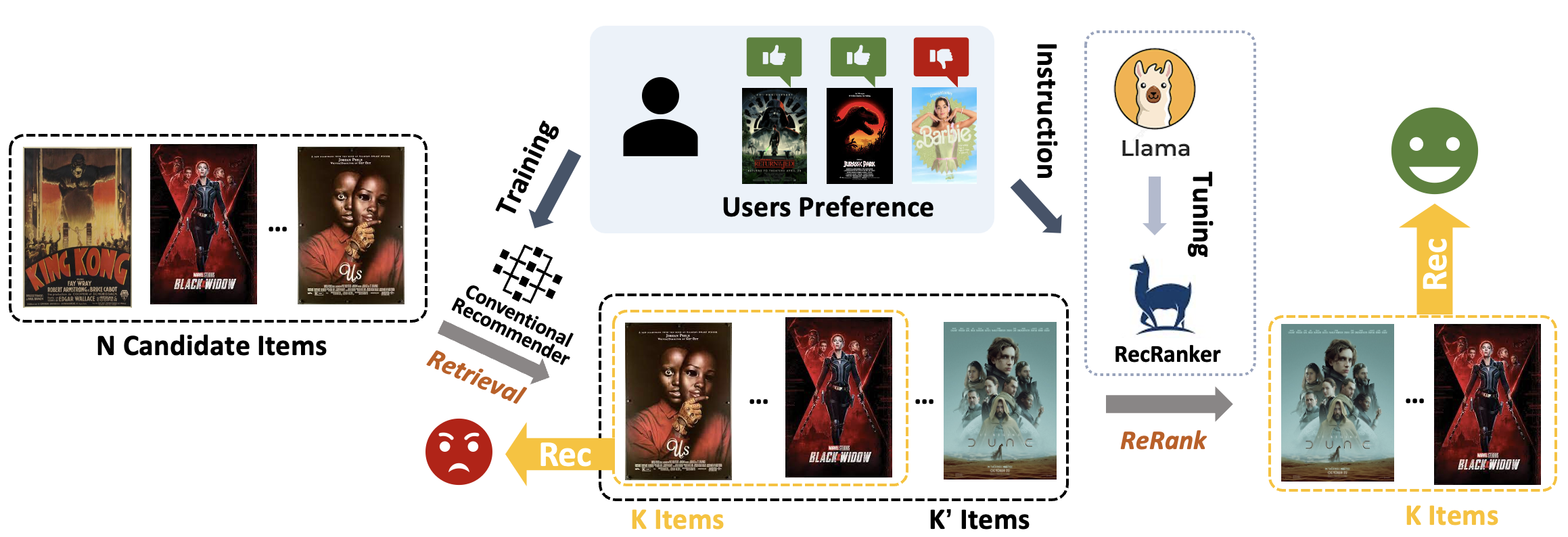}
    \caption{An example that illustrates the application of \model for the top-\textit{k} recommendation scenario.}
    \label{fig:mot}
    \vspace{-0.15in}
\end{figure*}

To address these shortfalls, we introduce instruction tuning {l}arge {la}nguage {m}odels {a}s \textbf{Ranker} for top-\textit{k} \textbf{Rec}ommendation, referred to as \textbf{\model}. Specifically, 
we propose an adaptive user sampling method to garner high-quality users, giving priority to users with a substantial history of interactions or who are representative of the broader user set, recognizing their heightened significance in the dataset.
To enhance the prompt, we propose a position shifting strategy to mitigate position bias.
In accordance with the concept of self-consistency in LLMs \cite{wang2022self}, we posit that the answer that receives consensus among most replies is more likely to be accurate.
We also incorporate signals from conventional recommendation models into prompts to augment LLM reasoning, as these signals can harness information from broader perspectives. 
The signals are seamlessly incorporated into the prompt using natural language descriptions in a uniform format.
Subsequently,
we curate an instruction-tuning dataset with enhanced prompts comprising three distinct ranking tasks, including pointwise, pairwise, and listwise ranking. 
The instruction tuning dataset is adopted to fine-tune the open-source LLM, resulting in a refined model that is well-aligned with the objectives of recommendation. 
Furthermore, we introduce a hybrid ranking approach that amalgamates all three ranking methods to bolster model performance. 
Experiments conducted on three real-world datasets validate the effectiveness of the proposed \model.
An example is shown in Figure \ref{fig:mot}, where \model is applied to rerank and improve the top-\textit{k} recommendation.


In a nutshell, our contribution is fourfold.
\begin{itemize}[leftmargin=*]


\item 
We introduce \model, a compact framework that applies instruction-tuned LLMs for diverse ranking tasks in top-\textit{k} recommendations. In addition, we propose a hybrid ranking method that ensembles various ranking tasks, aiming to further improve the model performance.
\item 
\model~employs adaptive user sampling to select high-quality users, thereby facilitating the construction of the instruction-tuning dataset. Furthermore, we propose a position shifting strategy within the prompt to mitigate the position bias in LLM.
\item 
Our approach incorporates information from conventional recommender systems into the instructions, enabling the LLM to synergistically leverage signals from both the conventional recommender system and textual information for better contextual understanding and user preferences reasoning.
\item 
We conduct extensive experiments on three real-world datasets to validate the effectiveness of our proposed \model. Impressively, 
\model~outperforms backbone models in most cases by a large margin, demonstrating its significant superiority.
\end{itemize}


\section{Related work}

In this section, we review related work, including the top-\textit{k} recommendations, LLMs as rankers in information retrieval, and LLMs in recommendation.

\subsection{Top-$k$ recommendation}
Top-$k$ recommendation \cite{yang2012top} has emerged as a burgeoning research field, aiming to suggest a list of $k$ items that are most likely to align with a user's preferences. Two predominant categories of algorithms for top-$k$ recommendation are collaborative filtering-based direct recommendation and sequential recommendation. For direct recommendation, 
memory-based approaches such as user-based and item-based collaborative filtering \cite{breese2013empirical} are proposed by leveraging the historical interactions between users and items to compute similarity scores and then generate recommendations. Advanced 
methods such as Neural Collaborative Filtering (NCF) \cite{he2017neural}, Neural Graph Collaborative Filtering (NGCF) \cite{wang2019neural}, and Lightweight Graph Convolution Network (LightGCN) \cite{he2020lightgcn} have been developed to better model collaborative user behavior and infer user preferences with more complex model structures. On the other hand, sequential recommendation emphasizes capturing users' dynamic behavior. Methods such as Gated Recurrent Unit for Recommendation (GRU4Rec) \cite{hidasi2015session}, Self-Attention-based Sequential Recommendation (SASRec) \cite{kang2018self}, and the recent bidirectional transformer-based BERT4Rec~\cite{sun2019bert4rec} leverage the sequentiality of user interactions to predict future items of interest.

Though conventional algorithms achieve promising results in top-\textit{k} recommendations, they still lack the ability to understand the content of the items. To address this issue, 
{this paper proposes to facilitate recommender systems by leveraging the contextual understanding and reasoning capabilities of LLMs.}

\begin{figure*}[]
    \centering
\includegraphics[width=\linewidth]{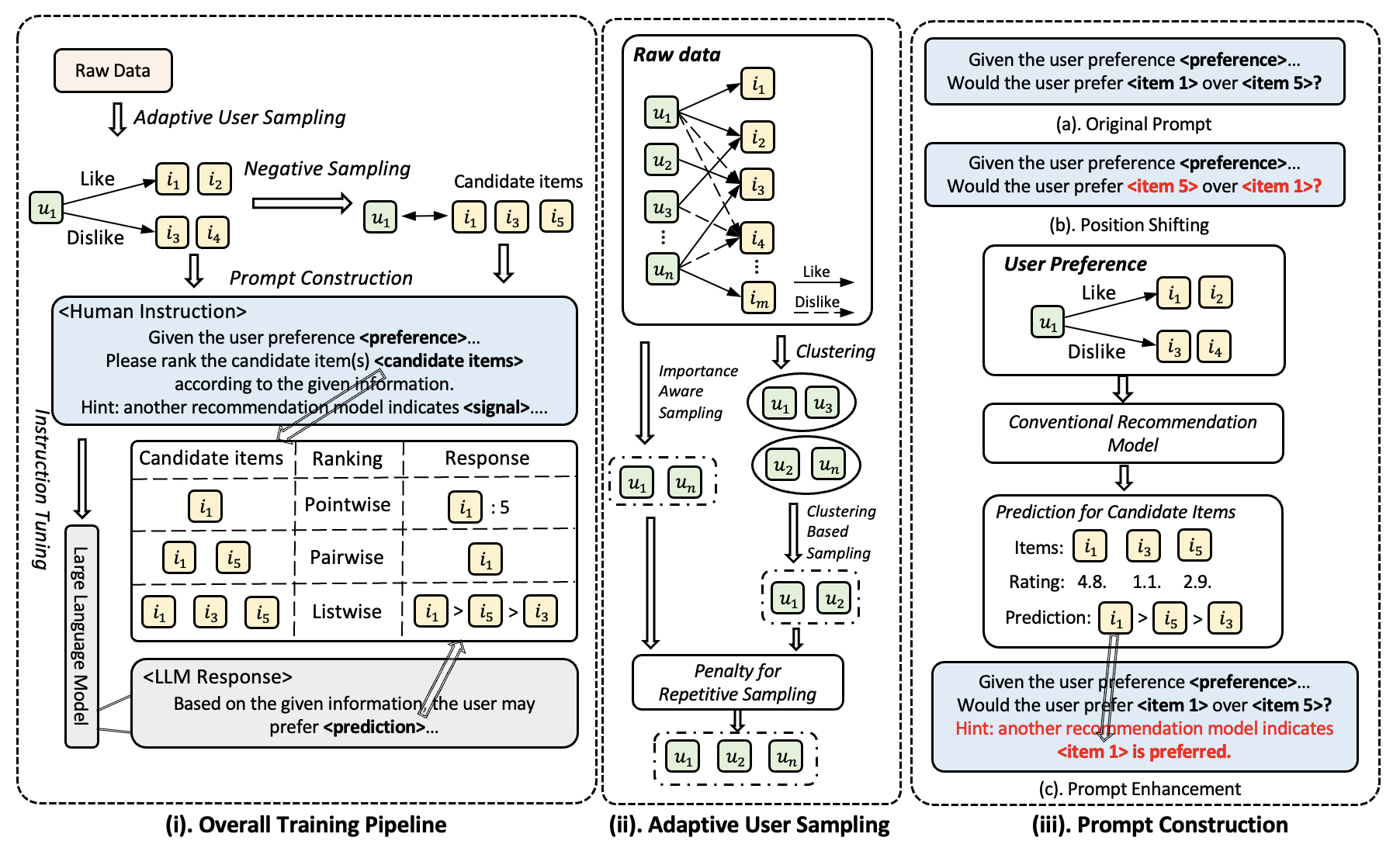}
    
    \caption{(i). The overall training pipeline of \model. (ii). Adaptive user sampling module, where we propose importance-aware sampling, clustering-based, and penalty for repetitive sampling to sample users. For each sampled user, corresponding candidate items are randomly selected from the items the user liked, disliked, and had no interaction with. 
    (iii). Prompt construction, where we incorporate position shifting and prompt enhancement strategies to enhance the model performance.  
    }
\label{fig:main}
\end{figure*}

\subsection{LLMs as Rankers in Information Retrieval}
There has been a strong recent interest in exploring using LLMs as rankers in information retrieval studies \cite{ma2023zero,qin2023large,sun2023chatgpt,pradeep2023rankvicuna,zhuang2023setwise,khramtsova2024leveraging}. For instance,  the Listwise Reranker with a Large Language Model (LRL) \cite{ma2023zero} demonstrates robust reranking efficiency without needing task-specific training data. It surpasses zero-shot pointwise methods and enhances the top-ranked results of a pointwise method. Similarly, the study by \cite{qin2023large} proposes Pairwise Ranking Prompting (PRP), a technique that lightens the burden on LLMs for document ranking. PRP could improve efficiency and compatibility with both generation and scoring LLM APIs. Furthermore, \cite{sun2023chatgpt} investigates the application of LLMs for relevance ranking in Information Retrieval and introduces a new test set NovelEval. Unlike these prompt engineering works, RankVicuna \cite{pradeep2023rankvicuna} is the first fully open-source LLM that addresses reproducibility and non-determinism in information retrieval reranking. It achieves comparable effectiveness to zero-shot reranking with GPT-3.5 using a smaller 7B parameter model.

Despite these advancements, the application of reranking in LLM-based recommendation tasks remains relatively less explored. Unlike text-rich scenarios (\textit{e.g.}, document ranking) in information retrieval, recommendation tasks often lack abundant text and require specialized processing. 

\subsection{LLMs for Recommendation}

Recently, LLMs have demonstrated remarkable capabilities and have found extensive applications across various domains, including recommender systems \cite{fan2023recommender,lin2023can}. 
Some recent works utilize LLMs for data augmentation \cite{wei2023llmrec,10447514}, generative recommendation \cite{li2023large,ji2024genrec}, or representation learning \cite{zhang2023collm,ren2023representation,yang2023large} in recommendations.
Notably, one strand of research leverages LLMs as rankers for recommender systems \cite{wang2023zero,zhang2023recommendation}. This approach is necessitated by the limitations of LLMs' fixed window size, which prevents the direct input of an exhaustive set of candidate items. Consequently, a retrieval model is commonly employed to refine and reduce the candidate item set. Specifically, Wang et al. \cite{wang2023zero} investigate the in-context-learning ability of LLMs with designed task-specific prompts to facilitate ranking tasks in sequential recommendation.
However, the misalignment between general-purpose LLMs and specialized recommendation tasks constrains the models' performance. To address this limitation, InstructRec \cite{zhang2023recommendation} instruction tunes LLMs using a specially constructed dataset of natural language instructions. However, existing research has yet to fully exploit the ranking capabilities of LLMs; it has primarily focused on singular ranking tasks, thereby leaving the ensemble of ranking tasks for improved performance largely unexplored.

To bridge this gap, we conduct a systematic investigation into the application of instruction-tuned LLMs for a variety of ranking tasks, including pointwise, pairwise, listwise, and their hybrid approaches, with the objective of fully elucidating the potential of LLMs in top-\textit{k} recommendation scenarios.

\section{PRELIMINARIES}

We consider a recommender system with a set of users, denoted $\mathcal{U} = \{ u_1, u_2, \ldots, u_n \}$, and a set of items, denoted $\mathcal{I} = \{ i_1, i_2, \ldots, i_m \}$. 
The top-$k$ recommendation focuses on identifying a subset of $k$ items $\mathcal{R}_u \subset \mathcal{I}$ for each user $u \in \mathcal{U}$. The subset is chosen to maximize a user-specific utility $U(u,\mathcal{R}_u)$ with the constraint $|\mathcal{R}_u| = k$, which is formally expressed as
\begin{equation}
 \mathcal{R}_u = {\arg\max}_{\mathcal{R}_u \subset \mathcal{I}, |\mathcal{R}_u| = k} U(u, \mathcal{R}_u). 
\end{equation}


Existing LLM-based recommendation methods typically
first utilize \textit{prompts} to interpret the recommendation task for user $u$ into natural language.
Given a prompt $\texttt{Prompt}_u$, the LLM-based recommendation for user $u$ with in-context learning is denoted by $\mathcal{R}_u = \texttt{LLM}(\texttt{Prompt}_u)$. 
To fine-tune our LLM using instruction-based approaches, we utilize a dedicated dataset, $\mathcal{D}_{ins}$. The resulting instruction-tuned LLM is represented as $\texttt{LLM}'$. Therefore, the recommendation process in the fine-tuned model can be represented as $\mathcal{R}_u = \texttt{LLM}'(\texttt{Prompt}_u)$.

\begin{table}
\caption{Table of Notations.}  
\begin{tabular}{l|l}  
\toprule 
\textbf{Notation} & \textbf{Definition} \\
\midrule
$\mathcal{U};\mathcal{I}$ & Set of users; Set of items. \\
$\mathcal{U}_1;\mathcal{U}_2;\mathcal{U}_3;\mathcal{U}_{ins}$ & Multiset of sampled users. \\
$\mathcal{D}_{ins}$ & Instruction tuning dataset. \\
$\mathcal{R}_u$ & Recommendation list for user $u$. \\
$\Theta$ & LLM parameter. \\
$k$ & The number of recommendation list. \\
$k'$ & The number of retrieval candidate items. \\
$U_{pointwise}$ & Utility score for pointwise ranking. \\
$U_{pairwise}$ & Utility score for pairwise ranking. \\
$U_{listwise}$ & Utility score for listwise ranking. \\
$U_{hybrid}$ & Utility score for hybrid ranking. \\
$\alpha_1, \alpha_2, \alpha_3$ & The hyper-parameters controlling utility scores. \\
\bottomrule 
\end{tabular}
\label{tab:notation}
\end{table}

\section{Methodology}

In this section, we first present the overall training and inference framework of our proposed \model paradigm
and then detail how we implement each component. Table \ref{tab:notation} summarizes the main
symbols and notations used in this work.

\subsection{Overview}

\begin{figure*}[]

    \centering
\includegraphics[width=1\linewidth]{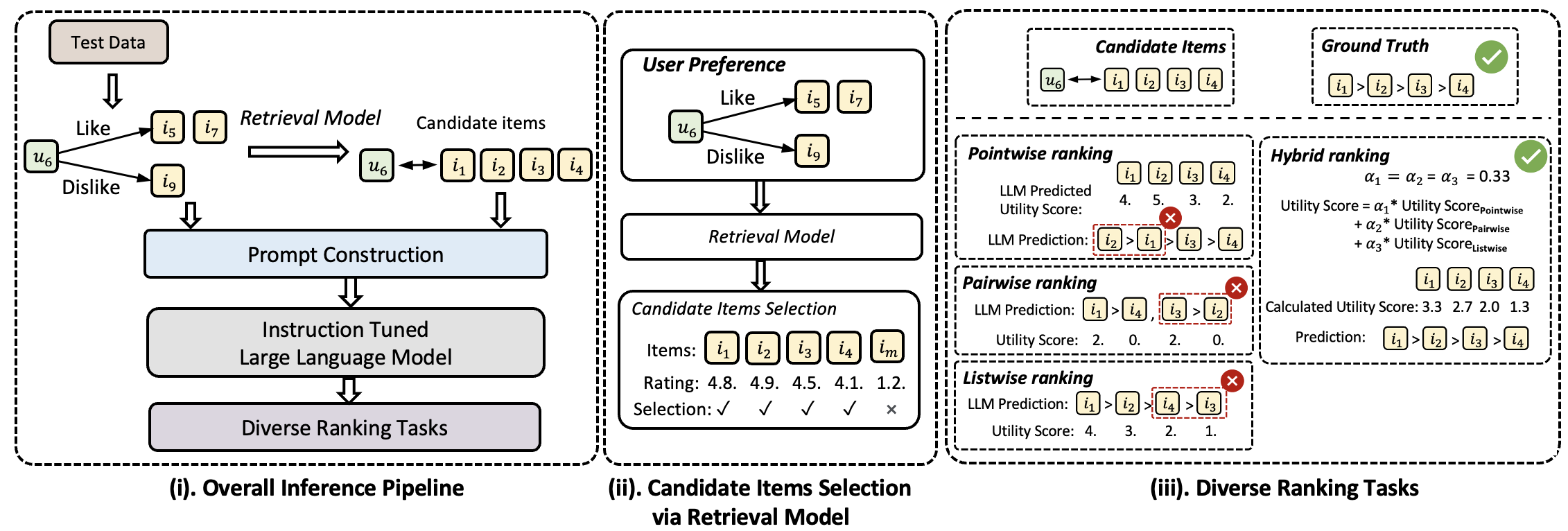}
    \caption{
    (i). The overall inference pipeline of \model. (ii). Candidate item selection via retrieval model, where we adopt the retrieval model to calculate the score for each item and select the highest ones as the candidate items.
    (iii). Comparison of the proposed hybrid ranking method with three ranking tasks during the inference stage.
    }
\label{fig:hybird}
\end{figure*}

The overall training and inference pipelines are depicted in Figure~\ref{fig:main} and Figure~\ref{fig:hybird}, respectively.
The training phase consists of four main stages: adaptive user sampling, candidate item selection via negative sampling, prompt construction, and instruction tuning. 
The adaptive user sampling stage aims to procure high-quality, representative, and diverse users. It incorporates three sampling strategies: importance-aware sampling, clustering-based sampling, and penalty for repetition.
For each user sampled, the candidate items consist of items liked and disliked by the users, as well as some un-interacted items selected via a commonly used negative sampling method \cite{rendle2012bpr,yang2020mixed}.
Given the users sampled and items selected, we construct prompts for each ranking task, augmenting them with signals from conventional recommender models. This strategy synergizes the strengths of both conventional recommender systems and textual data, thereby enhancing the system's overall performance.
Finally, we use the constructed data to fine-tune LLMs via instruction tuning. 

During the inference phase, for a user in the test data, we first select candidate items through a retrieval model. This item selection process is different from the training phase, where negative sampling is used. Subsequently, the prompt is constructed, following the same approach in the training phase. After that, the instruction-tuned LLM performs a variety of ranking tasks. Finally, a hybrid ranking method, which is achieved through the ensemble of multiple ranking tasks, is employed in this stage to enhance the model performance. 




\subsection{Adaptive User Sampling}
We first describe how we sample the raw recommendation dataset to create a list of users to be included in the fine-tuning dataset $\mathcal{D}_{ins}$.
We do not use the original user set $\mathcal{U}$, because we prefer to generate a list of users with improved quality and diversity. We denote such a list of users by a \textit{multiset} $\mathcal{U}_{ins}$. A multiset is a modified set that allows for multiple instances of the same element \cite{hein2003discrete}. A multiset is formally defined by a tuple $\mathcal{U}_{ins} = (\underline{\mathcal{U}}_{ins}, M_{ins})$, where $\underline{\mathcal{U}}_{ins}$ is the \textit{underlying set} of the multiset, consisting of its distinct elements, and $M_{ins}:\underline{\mathcal{U}}_{ins} \rightarrow \mathbb{Z}^+$ is the \textit{multiplicity} function, giving the number of occurrences of element $u \in \underline{\mathcal{U}}_{ins}$ as $M_{ins}(u)$. Therefore, the multiplicity $M_{ins}(u)$ of user $u$ will be the number of prompts regarding user $u$ in the instruction-tuning dataset $\mathcal{D}_{ins}$.

Some works sample users with equal probabilities from the user set $\mathcal{U}$ \cite{zhang2023instruction}, while other works sample nearest interactions \cite{bao2023tallrec}.
However, these methods could be sub-optimal, since the recommendation dataset often follows a long-tail distribution.
To compile a high-quality, representative, and diverse dataset, we introduce three {strategies:}
importance-aware sampling, clustering-based sampling, and penalty for repetitive sampling.
Specifically, we utilize importance-aware sampling and clustering-based sampling to create two multisets of candidate users, denoted by $\mathcal{U}_1$ and $\mathcal{U}_2$. Then from the combined multiset $\mathcal{U}_3 = \mathcal{U}_1 + \mathcal{U}_2$ with multiplicity function is $M_3 = M_1 + M_2$, we apply a penalty for repetitive sampling to select the final multiset $\mathcal{U}_{ins}$.

\subsubsection{Importance-aware Sampling}

Data in recommendation scenarios often exhibit a long-tail distribution, where a large number of items or users have minimal interactions, and a few have a large number of interactions \cite{luo2023improving,park2008long}. To optimize the quality of the data for building effective recommendation models, we propose an importance-aware sampling strategy. This {strategy}
prioritizes sampling from users with more interactions, based on the premise that users with a higher number of interactions provide more reliable and consistent data, crucial for modeling user preferences accurately.
We define 
the \textit{importance} of a user by the natural logarithm of their interaction count. The importance $w_u$ of user \(u\) is defined as $w_u = \ln(q_u)$,
where \(q_u\) denotes the number of interactions for user \(u\). The logarithmic scale is deliberately chosen to moderate the influence of users with extremely high interaction counts, ensuring that while they are given priority, they do not predominate the entire dataset.


The probability of selecting user $u$ is proportional to the importance $w_u$. This ensures that users with more interactions have a higher chance of being sampled, while still allowing for representation across the entire user base. In importance-aware sampling, the probability of sampling user \(u\) is
\begin{equation}
p_{u,\text{importance}} = \frac{w_u}{\sum_{v \in \mathcal{U}} w_v},
\end{equation}
where the denominator is the sum of the importance across all the users, serving as a normalizing factor so that the probabilities sum up to 1. Then $\mathcal{U}_1$ is denoted as
\begin{equation}
\mathcal{U}_1 = \{u: u \in \mathcal{U}\}\ \quad \text{with} \quad u \sim p_{u,\text{importance}}, 
\label{eq1}
\end{equation}
where $u \sim p_{u,importance}$ means the user $u$ is sampled with probability $p_{u,importance}$.

\subsubsection{Clustering-based Sampling}

To obtain representative users, we also employ a clustering-based sampling strategy. This {strategy} is grounded in the understanding that users in recommender systems exhibit diverse interests. By clustering users in the latent space, we can categorize them into distinct groups, each representing a unique set of interests. Such clustering enables us to capture the multifaceted nature of user preferences, ensuring that our sampling is not only representative but also encompasses the broad spectrum of user behaviors and tendencies.


Our framework allows for any clustering method, including but not limited to K-means \cite{hartigan1979algorithm}, Mean Shift \cite{comaniciu2002mean}, among others.
In this paper, we choose the K-means algorithm due to its effectiveness and simplicity in grouping data into cohesive clusters. We first represent each user as an \textit{embedding vector} derived by the retrieval model (\textit{e.g.}, LightGCN \cite{he2020lightgcn}), and then cluster the users into \( K \) groups based on the embedding vectors. We denote user \( u \)'s cluster by \( k_u \in \{1,\ldots,K\} \).
Once the users are clustered, we select samples from each cluster. This selection is not uniform but proportional to the size of each cluster. 
Mathematically, the sampling probability  of user $u$ in clustering-based sampling satisfies
\begin{equation}
p_{u,\text{clustering}} \propto \left|\left\{v \in \mathcal{U}:~k_v=k_u\right\}\right|,
\end{equation}
where $\left|\left\{v \in \mathcal{U}:~k_v=k_u\right\}\right|$ is the number of users in the same cluster as user $u$.
Similarity, the $\mathcal{U}_2$ is denoted as
\begin{equation}
\mathcal{U}_2 = \{u: u \in \mathcal{U}\}\ \quad \text{with} \quad u \sim p_{u,\text{clustering}}.
\label{eq2}
\end{equation}
This {strategy} not only preserves the diversity within each cluster but also ensures that larger clusters, which potentially represent more prevalent interests, have a proportionally larger representation in the final sample, thus ensuring a comprehensive representation of diverse user preferences.


\subsubsection{Penalty for Repetitive Sampling}

Given the two multisets $\mathcal{U}_1$ and $\mathcal{U}_2$ resulting from the importance-aware and clustering-based samplings, we need to construct the final user list $\mathcal{U}_{ins}$ from their sum $\mathcal{U}_3 = \mathcal{U}_1 + \mathcal{U}_2$, where the multiplicity function is $M_3 = M_1+M_2$.

To enhance diversity in the final multiset $\mathcal{U}_{ins}$, we implement a penalty for repetitive selections.
The rationale behind this {strategy} is to mitigate the overrepresentation of certain ``advantage groups'' –- users or items that might dominate the dataset due to their high frequency or popularity \cite{luo2023improving,park2008long}. To achieve this, we assign a penalty weight for each repeated selection within our sampling process. The penalty weight for a user $u \in \mathcal{U}_3$ is quantitatively expressed as \(\psi_u = C^{M_3(u)}\), where \(0 \textless  C \textless  1\) is a predefined constant. Thus, the penalty weight is decreasing in the number of occurrences  \(M_3(u)\).
This penalty weight directly influences the probability of a user being selected for the final dataset. To be specific, the probability of selecting user $u$ is 
\begin{equation}
p_{u,\text{penalty}} = \frac{\psi_u}{\sum_{v \in \mathcal{U}_3} \psi_{v}},    
\end{equation}
which ensures that those with higher occurrences are less likely to be chosen repeatedly.
Thus the $\mathcal{U}_{ins}$ is denoted as
\begin{equation}
\mathcal{U}_{ins} = \{u: u \in \mathcal{U}_3\}\ \quad \text{with} \quad u \sim p_{u,\text{penalty}}.
\label{eq3}
\end{equation}
This penalty for repetitiveness significantly enhances the diversity of the sample by reducing the likelihood of repeatedly selecting the same users and ensures a more equitable representation of less frequent users. 



\subsection{Candidate Items Selection}

The selection of candidate items differs between the training and inference phases. During training, negative sampling is utilized to select a mixture of items with which users have not interacted, as well as a random assortment of items that users have liked or disliked, forming the set of candidate items. While in the inference phase, a retrieval model is employed to generate the entire set of candidate items.

\subsubsection{Selection via Negative Sampling in the Training Phase}

In the training phase, the candidate item set includes randomly chosen items that users have liked and disliked. Besides, we employ the widely-used negative sampling technique \cite{rendle2012bpr,yang2020mixed,chen2023revisiting}, which involves randomly incorporating items with which users have not interacted into the candidate item set. These un-interacted items are considered as \textit{negative samples}. It is presumed that un-interacted items are more likely to be preferred over items that users have explicitly disliked.
Based on these selections, we establish the relative ranking comparison for the instruction tuning dataset construction.

\subsubsection{Selection via Retrieval Model in the Inference Phase}





In the realm of industrial recommender systems, platforms like YouTube often adopt a two-step process, initially utilizing a retrieval model to select a preliminary set of candidate items, which are subsequently re-ranked for final recommendations \cite{covington2016deep}. Specifically, within LLM-based recommender systems, the retrieval model plays a crucial role as a primary filter, effectively narrowing the scope of potential recommendations. This is particularly important due to the intrinsic limitations in the window size of LLMs. The architecture of the retrieval model is tailored to suit the nature of the recommendation task at hand. 
For direct recommendation,
models such as NCF \cite{he2017neural}, NGCF \cite{wang2019neural}, and LightGCN \cite{he2020lightgcn} are often employed. For sequential recommendation tasks, where the order of interactions is significant, models like SASRec \cite{kang2018self} and BERT4Rec \cite{sun2019bert4rec} are typically favored.

{
In the procedure of candidate item selection in the inference phase, we employ the retrieval model to compute a utility score for each item. Subsequently, we rank all the items based on their utility scores and select the top $k^\prime$ items with the highest scores as the candidate items.
For top-$k$ recommendation, 
this process will sample $k^\prime$ items with $k^\prime\textgreater k$.
}

\subsection{Prompt Construction}
\label{sec44}
In this section, we describe the construction of prompts. We begin by introducing a variety of ranking tasks, followed by a discussion of our proposed prompt enhancement method. This method involves augmenting prompts with signals from a conventional recommendation model.

\subsubsection{Pointwise, Pairwise, and Listwise Ranking}

Our recommender system incorporates a multifaceted approach to ranking tasks, encompassing pointwise, pairwise, and listwise rankings. Each of these methods plays a distinct role in evaluating and ordering candidate items based on their relevance to user preferences.
For 
pointwise ranking approach, each candidate item is assigned an individual relevance score. The entire list of candidates is then sorted based on these scores, providing a straightforward, score-based ranking.
The pairwise ranking method involves a direct comparison between two candidate items, determining which of the two is more relevant or preferable in a given context.
Differing from the above two, listwise ranking evaluates and sorts an entire list of candidate items. It considers the collective relevance of items, offering a comprehensive ranking based on overall suitability.
An example is {demonstrated}
in Table~\ref{tab:ex1}. Prompts are handcrafted in this paper.


Notably, in the vanilla setting, the time complexity (number of LLM calls) for pointwise, pairwise, and listwise ranking is $O(k')$, $O(k'^2)$, and $O(1)$, respectively. The pairwise ranking requires a substantial number of comparisons, as it needs to compare each pair from the list in the standard setting. To optimize the time complexity and reduce the cost, we intuitively found that if we change the ranking of two items either in the top $k$ of the list or after $k$ in this list, the evaluation metric hit ratio remains unchanged. Thus considering that the initial list derived by the retrieval model contains some form of prior knowledge, we randomly sample $k$ different pairs, where one pair consists of one item from the top $k$ positions and one from the positions after $k$. The intuition behind this is to replace the item in the top $k$ positions with items after the top $k$ positions, by utilizing the knowledge and ability of the LLM. In this way, the time complexity of pairwise ranking could be reduced to $O(k)$, which is much more acceptable.

\begin{table*}[t]
\caption{Illustrative examples of instructions for three ranking tasks. For better readability, a modified version of the actual instructions employed in our experiments is shown here.}
   \centering
   \resizebox{1\textwidth}{!}{
   \begin{tabular}{lll}
     \toprule
Type & Instructions & LLM Response \\\midrule
\multirow{2}{*}{Pointwise Ranking} & The historical interactions of a user include: \textless historical interactions\textgreater .& \multirow{2}{*}{5.} \\
&How would the user rate \textless candidate item\textgreater ?\\\midrule
\multirow{2}{*}{Pairwise Ranking} & The historical interactions of a user include: \textless historical interactions\textgreater . & \multirow{2}{*}{Yes.} \\
&Would the user prefer \textless candidate item 1\textgreater\ over \textless candidate item 2\textgreater ?\\\midrule
\multirow{2}{*}{Listwise Ranking}  & The historical interactions of a user include: \textless historical interactions\textgreater .& \multirow{2}{*}{\textless candidate item 1\textgreater,...,\textless candidate item 2\textgreater}\\
&How would the user rank the \textless candidate item list\textgreater ?
\\
\bottomrule
\end{tabular}\label{tab:ex1} }
\end{table*}

\subsubsection{Position Shifting in Prompt}
Position bias in LLMs arises when these models disproportionately favor items due to their locations in a list, rather than their inherent relevance or quality \cite{wang2023large,zheng2023judging}. This bias can significantly undermine the consistency and reliability of the output of the model. To mitigate the position bias, we adopt a position shifting strategy. During the training phase, we randomize the order of candidates and user preference items. This strategy is designed to prevent the model from prioritizing the item position over its actual significance. Similarly, in the inference phase, we continue this strategy by randomly altering the positions of the items. 
The primary objective of this strategy is to preserve those responses from LLMs that demonstrate consistency irrespective of item position. Consequently, the items identified are reflective of the model's true preferences, less influenced by position bias. By employing this method, we ensure that the LLMs’ responses are founded on genuine relevance, thereby enhancing the overall trustworthiness of the inference process.




\subsubsection{ Prompt Enhancement}
Existing LLM-based approaches often rely solely on LLMs for processing and ranking textual information. This reliance, however, neglects the rich and valuable signals that conventional recommendation models, like collaborative filtering, can offer. Models such as LightGCN \cite{he2020lightgcn} excel in extracting high-order collaborative signals, which play a pivotal role in understanding user preferences through the influences of user networks. The absence of the collaborative information could lead to less effective outcomes in LLM-based recommendations.

To bridge this gap, we propose a prompt enhancement method that integrates signals from conventional recommendation models into the prompts used for ranking tasks. This integration allows us to leverage the strengths of both LLMs and traditional recommendation models, creating a more informed and context-rich basis for decision-making.
Specifically, 
for pointwise ranking, we could utilize a rating prediction model like MF \cite{koren2009matrix} to forecast individual scores. These predictions are then transformed into natural language descriptions and seamlessly integrated into the prompt, providing a more nuanced basis for item evaluation.
For pairwise and listwise rankings, task-specific models such as LightGCN \cite{he2020lightgcn} and SASRec \cite{kang2018self} are employed to predict rankings. 
In this paper, we adopt MF \cite{koren2009matrix} and the LightGCN \cite{he2020lightgcn} model for prompt enhancement.
The insights from these predictions are then incorporated into the prompts, enhancing the context and depth of the ranking process.
By augmenting prompts with data from conventional recommendation models, our method significantly enriches the ranking tasks in recommender systems. This innovative approach not only capitalizes on the advanced capabilities of LLMs but also harnesses the collaborative or sequential information offered by conventional recommendation models. 


\subsection{ Optimization via Instruction Tuning }
After constructing the dataset, we focus on fine-tuning the LLM in a supervised manner, specifically through instruction tuning. This process involves optimizing the LLM using a dataset generated from instructional data, aligning the model responses more closely with user intents and preferences.


The approach we adopt for supervised fine-tuning is grounded in the standard cross-entropy loss, following the principles outlined in Alpaca \cite{alpaca}. The core of this process lies in the training set \( \mathcal{D}_{ins} \), which is comprised of natural language instruction input-output pairs \((x, y)\). This dataset is instrumental in guiding the fine-tuning process, ensuring that the model outputs are aligned with the structured instructional data.


The primary objective in this phase is to fine-tune the pre-trained LLM by minimizing the cross-entropy loss. This is mathematically formalized as:
\begin{equation}
\min _{\Theta} \sum_{(x, y) \in \mathcal{D}_{ins}} \sum_{t=1}^{|y|} -\log P_{\Theta}\left(y_{t} \mid x, y_{[1:t-1]}\right),
\label{eq4}
\end{equation}
where \( \Theta \) represents the model parameters,  \( P_{\Theta} \) denotes the conditional probability of generating the \( t \)-th token \( y_t \) in the target output \( y \), given the input \( x \) and the preceding tokens \( y_{[1:t-1]} \), and \( |y| \) is the length of the target sequence \( y \).


By minimizing this loss function, the model parameters \( \Theta \) are fine-tuned to better align with instructional tuning dataset \( \mathcal{D}_{ins} \). This fine-tuning leverages the LLM's pre-existing capabilities in general language understanding and reasoning, as acquired during its initial training phase. 
This results in a refined model adept at capturing and interpreting user preferences articulated in natural language, crucial for succeeding recommendation tasks. Consequently, the LLM can provide recommendations more attuned to the user's personalized needs and preferences, enhancing the system's effectiveness.
The training process of \model is shown in Algorithm \ref{alg:train}.

\begin{algorithm}[tb]
\SetKwData{False}{False}\SetKwData{This}{this}\SetKwData{Up}{up}
\SetKwFunction{Union}{Union}
\SetKwComment{tcc}{/*\ }{}
\SetKwFunction{server}{ServerRun}
\SetKwFunction{AdaptiveSampling}{AdaptiveSampling}
\SetKwInOut{Input}{Input}\SetKwInOut{Output}{Output}
\SetKwProg{Fn}{Function}{:}{\KwRet}

\Input{\ Original LLM parameter $\Theta$, user set $\mathcal{U}$, item set $\mathcal{I}$ 
}

\Output{\ Instruction tuned LLM parameter $\Theta'$}
\BlankLine

$\mathcal{U}_{ins} \leftarrow$ Sampling user according to \AdaptiveSampling{$\mathcal{U}$}\;

$\mathcal{D}_{ins} \leftarrow$ Prompt construction and enhancement for $\mathcal{U}_{ins}$ and $\mathcal{I}$ according to Sec.~\ref{sec44}\;
\While{not converge}{
 Calculate the loss and optimize the model parameters $\Theta$ according to Eq. \eqref{eq4}
}
\textbf{return $\Theta'$}\;

\Fn{\AdaptiveSampling{$\mathcal{U}$}}{
$\mathcal{U}_1 \leftarrow$ Perform importance-aware sampling for $\mathcal{U}$ according to Eq. \eqref{eq1}\;
$\mathcal{U}_2 \leftarrow$ Perform clustering-based sampling for $\mathcal{U}$  according to Eq. \eqref{eq2}\;
$\mathcal{U}_{ins} \leftarrow$ Apply repetitive sampling penalty for $\mathcal{U}_3=\mathcal{U}_1 + \mathcal{U}_2$ according to Eq. \eqref{eq3}\;
\KwRet $\mathcal{U}_{ins}$;
}

\caption{Training Procedure of \model}
\label{alg:train}
\end{algorithm}



\subsection{Hybrid Ranking}

LLM-based recommendation with singular ranking method may have some limitations: (1) the performance of different ranking methods can vary significantly across various recommendation scenarios, which complicates the selection of an appropriate ranking method; (2) LLMs may produce inconsistent results due to their inherent randomness mechanism \cite{hinton2012practical,holtzman2019curious} or variations in prompts~\cite{gu2023systematic,wei2022chain}, resulting in sub-optimal model performance.

Recent research highlights the self-consistency in LLM \cite{wang2022self}, suggesting that the result agreed by most LLM responses has a higher probability of being correct. Inspired by this, it could be promising to combine the LLM response for different ranking prompts to obtain more promising recommendation performance. 
However, the amalgamation of these responses is not a straightforward process. Different LLM responses may not be directly compatible, necessitating the development of effective strategies to facilitate their combination.

To address the aforementioned issues, we propose a hybrid ranking approach to align different LLM responses within a unified space. We achieve this by assigning a \textit{utility score} for items in each LLM response generated by different ranking methods. For instance, when the LLM predicts that a certain item is more likely to be preferred, a higher utility score is assigned to that item. This principle is applicable across all ranking methods, thereby enabling an aggregation of LLM responses.

More specifically, we assign a corresponding utility score for each ranking task.
For the pointwise ranking task, the utility score, \(U_{pointwise}\), is initially determined by the relevance score from the LLM prediction. To refine this score and differentiate between items with identical ratings, an additional utility score from the retrieval model is incorporated. This score is denoted as \(U_{retrieval} = -m \cdot \mathcal{C}_1\), where \(\mathcal{C}_1\) is a constant and \(m\), representing the item's position as determined by the retrieval model, varies from 1 to \(k^\prime\) (\textit{i.e.}, the total number of candidate items). Therefore, the comprehensive utility score for the pointwise ranking task is \(U_{pointwise} = U_{retrieval} + \mathcal{P}\), where $\mathcal{P}$ is the LLM prediction.
In the pairwise ranking scenario, items preferred by the LLM are attributed a utility score \(U_{pairwise} = \mathcal{C}_2\), where \(\mathcal{C}_2\) is another constant.
For listwise ranking, the formula \(U_{listwise} = -m' \cdot \mathcal{C}_3\) is employed to score each item, with \(m'\) representing the position predicted by the LLM and varying from 1 to \(k^\prime\) and \(\mathcal{C}_3\) being a constant. This formula assigns scores across the list of items, integrating the listwise perspective into the hybrid approach.

The hybrid ranking method operates by amalgamating the utility scores derived from three distinct ranking tasks. While this process could involve complex reflection or even the use of LLMs to make the selection \cite{yue2023large,chen2023frugalgpt}, we opt for a simplified approach of direct linear interpolation of utility scores. Mathematically, this process can be expressed as:
\begin{equation}
U_{hybrid} = \alpha_1 U_{pointwise} + \alpha_2 U_{pairwise} + \alpha_3 U_{listwise}
\end{equation}
where \(\alpha_1\), \(\alpha_2\), and \(\alpha_3\) are weighting coefficients that sum up to 1. The values of these coefficients determine the hybrid ranking's ability to effectively emulate any of the individual ranking methods, thereby providing flexibility in the recommendation approach.
Recognizing that each ranking task encapsulates different facets of the recommendation problem, the hybrid ranking method aims to combine the strengths of each individual task for achieving a more comprehensive and effective recommendation process.

\section{Experiment}

The primary goal is to investigate the extent to which integrating the introduced model can improve the performance of current recommender systems. 
{Therefore, we conduct comprehensive experiments to answer the following research questions: }


\begin{itemize}[leftmargin=*]
\item
\textbf{RQ1}: Does the proposed \model~framework enhance the performance of existing recommendation models?
\item
\textbf{RQ2}: What is the impact of model efficiency and the reranking process of \model?
\item
\textbf{RQ3}: How do the key components of \model, such as adaptive sampling and prompt enhancement, contribute to the quality of recommendation respectively?
\item
\textbf{RQ4}: How do various hyper-parameters influence the overall performance of the framework?
\item
\textbf{RQ5}: How does the instruction-tuned model compare to off-the-shelf LLMs, such as the state-of-the-art GPT-4 model?
\end{itemize}

\subsection{Experimental Setup}



\subsubsection{Dataset}
Following \cite{bao2023tallrec}, we rigorously evaluate the performance of our proposed framework by employing three heterogeneous, real-world datasets. 
\textbf{MovieLens}\footnote{\url{https://grouplens.org/datasets/movielens/}} \cite{harper2015movielens} dataset is utilized as a standard benchmark in movie recommender systems. We explore two subsets of this dataset: MovieLens-100K (ML-100K), containing 100,000 user-item ratings, and MovieLens-1M (ML-1M), which expands to approximately 1 million ratings. \textbf{BookCrossing}\footnote{In the absence of timestamp data within the BookCrossing dataset, we have reconstructed historical interactions via random sampling.} \cite{ziegler2005improving} dataset comprises user-submitted book ratings on a 1 to 10 scale and includes metadata such as ‘Book-Author' and `Book-Title'.
We apply the 10-core setting \cite{chang2021sequential} to assure data quality, which involves excluding users and items that have fewer than ten interactions from the BookCrossing dataset. 
The key statistics of these datasets are detailed in Table~\ref{tab:dataset}.


\begin{table}[t]
\centering
\caption{Dataset Description.}
\label{tab:transposed_dataset}
\begin{tabular}{c c c c c}
\toprule
Dataset & \# of User & \# of Item & \# of Rating & Density \\
\midrule
ML-100K & 943 & 1,682 & 100,000 & 0.063046 \\
ML-1M & 6,040 & 3,706 & 1,000,209 & 0.044683 \\
BookCrossing & 1,820 & 2,030 & 41,456 & 0.011220 \\
\bottomrule
\end{tabular}\label{tab:dataset}
\end{table}

\subsubsection{Evaluation Metrics}
In line with the methodologies adopted in prior works \cite{he2020lightgcn,sun2019bert4rec}, we employ two well-established metrics for evaluating the top-$k$ recommendation task: Hit Ratio (HR) and Normalized Discounted Cumulative Gain (NDCG), denoted as H and N, respectively. Formally, they are defined by the following equations:
\begin{equation}
\label{eq:aug3}
    \begin{split}
    & \mathrm{H}@k = \frac{1}{|\mathcal{U}|} \sum_{u \in \mathcal{U}} \mathds{1} \left(R_{u,g_u} \le k\right),\\
    & \mathrm{N} @ k=\frac{1}{|\mathcal{U}|} \sum_{u \in \mathcal{U}}  \frac{2^{\mathds{1}\left(R_{u, g_{u}} \leq k\right)}-1}{\log _{2}\left(\mathds{1}\left(R_{u, g_{u}} \leq k\right)+1\right)},
    \end{split}
\end{equation}
where $\mathds{1}(\cdot)$ is the indicator function, $R_{u, g_{u}}$ 
is the rank predicted by the model for the
ground truth item $g_{u}$ and user $u$.

Our experimental setup involves setting $k$ to either 3 or 5, similar to the evaluation approach detailed in \cite{zhang2023recommendation}, allowing for a comprehensive assessment.

\subsubsection{Data Preprocessing}
We adopt the leave-one-out evaluation strategy, aligning with the methodologies employed in prior research \cite{zhang2023recommendation, luo2022hysage}. Under this strategy, the most recent interaction of each user is assigned as the test instance, the penultimate interaction is used for validation, and all preceding interactions constitute the training set.
Regarding the construction of the instruction-tuning dataset, 
we sampled 10,000 instructions for each ranking task for the ML-1M dataset.
In the case of the ML-100K and BookCrossing datasets, we formulated 5,000 instructions for each task, respectively.
We eliminated instructions that were repetitive or of low quality (identified by users with fewer than three interactions in their interaction history), leaving approximately 56,000 high-quality instructions.
These instructions are then combined to create a comprehensive instruction-tuning dataset, which is utilized to fine-tune the LLM. 


\subsubsection{Model Selection}


We incorporate our \model~with the following recommendation models as the backbone models: 


\begin{itemize}[leftmargin=*]
    \item \textbf{\textit{Direct Recommendation.}}  For direct recommendation, we select four widely recognized methods, namely:
\begin{itemize}
    \item \textbf{MF} \cite{koren2009matrix}: This is a classic collaborative filtering method that decomposes the user-item interaction matrix.
    \item \textbf{LightGCN} \cite{he2020lightgcn}: An advanced model that simplifies the design of graph convolution network (GCN) for recommendation, enhancing the performance by eliminating unnecessary transformations and nonlinearities in GCN.
    \item \textbf{MixGCF} \cite{huang2021mixgcf}: This model proposes a mixed collaborative filtering framework for recommendation, which can effectively alleviate the sparsity and cold-start problem.
    \item \textbf{SGL} \cite{wu2021self}: A novel graph learning model that captures the global structure of the graph via self-supervised learning.
\end{itemize}
\item \textbf{\textit{Sequential Recommendation.}} 
     For sequential recommendation, we adopt three representative models, including:
\begin{itemize}
    \item \textbf{SASRec} \cite{kang2018self}: This is the first method that utilizes the masked multi-head attention mechanism to model users’ sequential behaviors. 
    \item \textbf{BERT4Rec} \cite{sun2019bert4rec}: This model employs the BERT architecture, originally designed for natural language processing tasks, to capture the sequential patterns in users' interaction histories.
    \item \textbf{CL4SRec} \cite{xie2022contrastive}: A contrastive learning-based method for sequential recommendation, which captures the global temporal dynamics and local item-item transitions.
\end{itemize}
\end{itemize}



    

The backbone models serve as the retrieval models in \model.
For each backbone model, we choose the top ten items as candidate items, setting \( k^\prime = 10 \).





We leave out the \textbf{direct comparison} with other instruction-tuning LLM for recommendation methods such as TALLRec \cite{bao2023tallrec} and InstructRec \cite{zhang2023instruction}. This exclusion is justified as these methods are not primarily designed for diverse ranking tasks. Specifically, TALLRec is tailored for a binary classification task, determining whether a user likes an item or not. InstructRec, on the other hand, relies on the powerful yet closed-source GPT model to generate information, rendering it impractical in our context. Nevertheless, it is important to note that these methods adhere to the standard approach for instruction tuning in LLMs. As detailed in Section \ref{sec:ab}, we include an ablation study that evaluates our method's enhancements over the standard instruction tuning LLMs, thereby underscoring the superiority of our approach.

\begin{table*}[t]
\centering
\caption{ Performance achieved by different direct recommendation methods. The best results are highlighted in \textbf{boldfaces}. }
\label{tab:t2}
\begin{adjustbox}{width=\linewidth}
\begin{tabular}{clc c c ccccccccccc}
\toprule

\multirow{2}{*}{Backbone} & \multirow{2}{*}{Method} & \multicolumn{4}{c}{ML-100K}  & \multicolumn{4}{c}{ML-1M}& \multicolumn{4}{c}{BookCrossing}  \\ \cmidrule(lr){3-6}\cmidrule(lr){7-10}\cmidrule(lr){11-14}
\multirow{2}{*}{}&\multirow{2}{*}{}& H@3 $\uparrow$ &  N@3 $\uparrow$& H@5 $\uparrow$ &  N@5 $\uparrow$& H@3 $\uparrow$ &  N@3 $\uparrow$& H@5 $\uparrow$ &  N@5 $\uparrow$& H@3 $\uparrow$ &  N@3 $\uparrow$& H@5 $\uparrow$ &  N@5 $\uparrow$ \\ \midrule

\multirow{6}{*}{MF} & Base & 0.0455 & 0.0325 & 0.0690 & 0.0420 & 0.0255 & 0.0187 & 0.0403 & 0.0248 & 0.0503 & 0.0389 & 0.0689 & 0.0465\\ 

\multirow{6}{*}{}& \modell$_{pointwise}$ 
& 0.0660 & 0.0486 & 0.0917 & 0.0592
& 0.0294 & 0.0213 & 0.0456 & 0.0279 
& 0.0872 & 0.0710 & \textbf{0.0966} & 0.0749 
\\
\multirow{6}{*}{}& \modell$_{pairwise}$ 
& 0.0533 & 0.0368 & 0.0783 & 0.0471 
& 0.0275 & 0.0201 & 0.0438 & 0.0268
& 0.0539 & 0.0419 & 0.0716 & 0.0492
\\
\multirow{6}{*}{}& \modell$_{listwise}$ 
& 0.0464 & 0.0346 & 0.0712 & 0.0448 
& 0.0271 & 0.0196 & 0.0416 & 0.0256
& 0.0430 & 0.0312 & 0.0674 & 0.0411
\\
\multirow{6}{*}{}& \modell$_{hybrid}$ 
& \textbf{0.0690} & \textbf{0.0513} & \textbf{0.0919} & \textbf{0.0607}
& \textbf{0.0312} & \textbf{0.0230} & \textbf{0.0469} & \textbf{0.0294}
& \textbf{0.0873} & \textbf{0.0720} & \textbf{0.0966} & \textbf{0.0759}

\\
\multirow{6}{*}{}& Improvement & 51.65\%&57.85\%&33.19\%&44.52\%&22.35\%&22.99\%&16.38\%&18.55\%&73.56\%&85.09\%&40.20\%&63.23\%
\\\midrule
\multirow{6}{*}{LightGCN} & Base&0.0492&0.0343&0.0744&0.0447&0.0273&0.0197&0.0431&0.0261&0.0645&0.0499&0.0875&0.0595\\
\multirow{6}{*}{}& \modell$_{pointwise}$&0.0723&0.0524&\textbf{0.0990}&\textbf{0.0634}&\textbf{0.0324}&\textbf{0.0232}&0.0480&0.0296&0.1076&0.0876&\textbf{0.1231}&0.0940\\
\multirow{6}{*}{}& \modell$_{pairwise}$&0.0414&0.0298&0.0645&0.0393&0.0287&0.0205&0.0450&0.0272&0.0622&0.0481&0.0840&0.0572\\
\multirow{6}{*}{}& \modell$_{listwise}$&0.0509&0.0380&0.0764&0.0485&0.0281&0.0202&0.0440&0.0267&0.0566&0.0395&0.0854&0.0513\\
\multirow{6}{*}{}& \modell$_{hybrid}$&\textbf{0.0731}&\textbf{0.0527}&0.0971&0.0625&0.0320&\textbf{0.0232}&\textbf{0.0497}&\textbf{0.0305}&\textbf{0.1088}&\textbf{0.0888}&0.1219&\textbf{0.0942}\\

\multirow{6}{*}{}& Improvement & 48.58\%&53.64\%&33.06\%&41.83\%&18.68\%&17.77\%&15.31\%&16.86\%&68.68\%&77.96\%&40.69\%&58.32\%
\\\midrule
\multirow{6}{*}{MixGCF} & Base&0.0537&0.0412&0.0736&0.0492&0.0144&0.0108&0.0232&0.0144&0.0746&0.0584&0.0957&0.0671\\
\multirow{6}{*}{}& \modell$_{pointwise}$&0.0701&0.0542&0.0930&0.0637&0.0170&0.0126&0.0263&0.0164&\textbf{0.1113}&0.0916&0.1208&0.0955\\
\multirow{6}{*}{}& \modell$_{pairwise}$&0.0537&0.0413&0.0770&0.0508&0.0169&0.0123&0.0265&0.0163&0.0686&0.0556&0.0885&0.0638\\
\multirow{6}{*}{}& \modell$_{listwise}$&0.0507&0.0376&0.0738&0.0470&0.0154&0.0111&0.0252&0.0151&0.0558&0.0415&0.0823&0.0523\\
\multirow{6}{*}{}& \modell$_{hybrid}$&\textbf{0.0712}&\textbf{0.0551}&\textbf{0.0932}&\textbf{0.0641}&\textbf{0.0180}&\textbf{0.0133}&\textbf{0.0269}&\textbf{0.0169}&\textbf{0.1113}&\textbf{0.0918}&\textbf{0.1209}&\textbf{0.0958}\\

\multirow{6}{*}{}& Improvement & 32.59\%&33.74\%&26.63\%&30.28\%&25.00\%&23.15\%&15.95\%&17.36\%&49.20\%&57.19\%&26.33\%&42.77\%
\\\midrule
\multirow{6}{*}{SGL} & Base&0.0505&0.0380&0.0729&0.0472&0.0284&0.0206&0.0434&0.0267&0.0609&0.0476&0.0812&0.0560\\
\multirow{6}{*}{}& \modell$_{pointwise}$&\textbf{0.0693}&0.0517&\textbf{0.0885}&0.0596&0.0320&0.0230&0.0492&0.0301&\textbf{0.0951}&\textbf{0.0793}&0.1044&\textbf{0.0831}\\
\multirow{6}{*}{}& \modell$_{pairwise}$&0.0470&0.0349&0.0710&0.0447&0.0292&0.0211&0.0451&0.0275&0.0590&0.0468&0.0785&0.0549\\
\multirow{6}{*}{}& \modell$_{listwise}$&0.0535&0.0385&0.0740&0.0470&0.0286&0.0207&0.0436&0.0269&0.0493&0.0362&0.0760&0.0471\\
\multirow{6}{*}{}& \modell$_{hybrid}$&0.0690&\textbf{0.0525}&0.0882&\textbf{0.0604}&\textbf{0.0325}&\textbf{0.0235}&\textbf{0.0497}&\textbf{0.0305}&0.0950&0.0791&\textbf{0.1045}&\textbf{0.0831}\\
\multirow{6}{*}{}& Improvement & 37.23\%&38.16\%&21.40\%&27.97\%&14.44\%&14.08\%&14.52\%&14.23\%&56.16\%&66.60\%&28.69\%&48.39\%
\\

\bottomrule
\end{tabular}
\end{adjustbox}
\end{table*}

\subsubsection{Implementation Details}
We chose LLaMA-2 (7B) \cite{touvron2023llama2} as the backbone of LLM in our experiment due to its strong capability among the open-source LLMs.
In the training phase of LLaMA-2 (7B), we adopted a uniform learning rate of 
\(2 \times 10^{-5}\) , coupled with a context length of 1024. The batch size was fixed at 4, complemented by gradient accumulation steps of 2. Additionally, a cosine scheduler was implemented, integrating a preliminary warm-up phase of 50 steps. The training comprised a total of 6000 steps. We employed DeepSpeed's ZeRO-3 stage optimization \cite{rajbhandari2020zero} alongside the flash attention technique \cite{dao2022flashattention} for efficient training of these models. This training process was executed on 16 NVIDIA A800 80GB GPUs. During the inference process, the vLLM framework \cite{kwon2023efficient} was employed, setting the temperature parameter at 0.1, with top-$k$ and top-$p$ values at 10 and 0.1, respectively. Inference was conducted using a single NVIDIA A800 80GB GPU.

\begin{table*}[h]
\centering
\caption{ Performance achieved by different sequential recommendation methods. The best results are highlighted in \textbf{boldfaces}. }
\label{tab:t3}
\begin{adjustbox}{width=\linewidth}
\begin{tabular}{clc c c ccccccccccc}
\toprule

\multirow{2}{*}{Backbone} & \multirow{2}{*}{Method} & \multicolumn{4}{c}{ML-100K}  & \multicolumn{4}{c}{ML-1M}& \multicolumn{4}{c}{BookCrossing}  \\ \cmidrule(lr){3-6}\cmidrule(lr){7-10}\cmidrule(lr){11-14}
\multirow{2}{*}{}&\multirow{2}{*}{}& H@3 $\uparrow$ &  N@3 $\uparrow$& H@5 $\uparrow$ &  N@5 $\uparrow$& H@3 $\uparrow$ &  N@3 $\uparrow$& H@5 $\uparrow$ &  N@5 $\uparrow$& H@3 $\uparrow$ &  N@3 $\uparrow$& H@5 $\uparrow$ &  N@5 $\uparrow$ \\ \midrule

\multirow{6}{*}{SASRec} & Base&0.0187&0.0125&0.0385&0.0205&0.0277&0.0165&0.0501&0.0257&0.0150&0.0086&0.0279&0.0139\\
\multirow{6}{*}{}& \modell$_{pointwise}$&0.0206&0.0147&0.0432&0.0239&\textbf{0.0308}&0.0206&\textbf{0.0541}&0.0301&0.0351&0.0247&0.0482&0.0302\\
\multirow{6}{*}{}& \modell$_{pairwise}$&\textbf{0.0277}&\textbf{0.0190}&\textbf{0.0479}&\textbf{0.0273}&0.0289&0.0195&0.0502&0.0282&0.0227&0.0149&0.0357&0.0203\\
\multirow{6}{*}{}& \modell$_{listwise}$&0.0204&0.0149&0.0321&0.0197&0.0239&0.0162&0.0407&0.0231&0.0218&0.0153&0.0323&0.0196\\
\multirow{6}{*}{}& \modell$_{hybrid}$&0.0232&0.0160&0.0436&0.0243&0.0304&\textbf{0.0212}&0.0526&\textbf{0.0303}&\textbf{0.0381}&\textbf{0.0270}&\textbf{0.0487}&\textbf{0.0315}
\\
\multirow{6}{*}{}& Improvement & 48.13\%&52.00\%&24.42\%&33.17\%&11.19\%&28.48\%&7.98\%&17.90\%&154.00\%&213.95\%&74.55\%&126.62\%
\\\midrule

\multirow{6}{*}{BERT4Rec} & Base&0.0153&0.0104&0.0294&0.0161&0.0107&0.0069&0.0211&0.0112&0.0179&0.0119&0.0343&0.0185\\
\multirow{6}{*}{}& \modell$_{pointwise}$&0.0183&0.0129&0.0334&0.0191&\textbf{0.0140}&\textbf{0.0095}&\textbf{0.0231}&\textbf{0.0133}&0.0390&0.0279&0.0557&0.0348\\
\multirow{6}{*}{}& \modell$_{pairwise}$&\textbf{0.0194}&\textbf{0.0133}&0.0334&0.0190&0.0090&0.0061&0.0159&0.0089&0.0254&0.0173&0.0416&0.0240\\
\multirow{6}{*}{}& \modell$_{listwise}$&0.0162&0.0119&0.0240&0.0151&0.0124&0.0085&\textbf{0.0231}&0.0128&0.0242&0.0168&0.0371&0.0221\\
\multirow{6}{*}{}& \modell$_{hybrid}$&0.0191&0.0130&\textbf{0.0343}&\textbf{0.0192}&0.0135&0.0094&0.0230&\textbf{0.0133}&\textbf{0.0422}&\textbf{0.0305}&\textbf{0.0566}&\textbf{0.0365}
\\
\multirow{6}{*}{}& Improvement & 26.80\%&27.88\%&16.67\%&19.25\%&30.84\%&37.68\%&9.48\%&18.75\%&135.75\%&156.30\%&65.01\%&97.30\%
\\\midrule

\multirow{6}{*}{CL4SRec} & Base&\textbf{0.0243}&0.0143&\textbf{0.0436}&0.0222&0.0259&0.0153&0.0492&0.0248&0.0151&0.0088&0.0282&0.0141\\
\multirow{6}{*}{}& \modell$_{pointwise}$&0.0219&0.0149&0.0417&\textbf{0.0229}&\textbf{0.0290}&\textbf{0.0196}&\textbf{0.0519}&\textbf{0.0289}&0.0349&0.0241&0.0501&0.0304\\
\multirow{6}{*}{}& \modell$_{pairwise}$&0.0226&\textbf{0.0155}&0.0404&\textbf{0.0229}&0.0278&0.0186&0.0462&0.0261&0.0190&0.0132&0.0361&0.0202\\
\multirow{6}{*}{}& \modell$_{listwise}$&0.0200&0.0145&0.0323&0.0195&0.0232&0.0156&0.0413&0.0230&0.0219&0.0152&0.0323&0.0194\\
\multirow{6}{*}{}& \modell$_{hybrid}$&0.0221&0.0152&0.0400&0.0224&0.0280&0.0192&0.0512&0.0286&\textbf{0.0375}&\textbf{0.0263}&\textbf{0.0514}&\textbf{0.0321}\\
\multirow{6}{*}{}& Improvement & N/A&8.39\%&N/A&3.15\%&11.97\%&28.10\%&5.49\%&16.53\%&148.34\%&198.86\%&82.27\%&127.66\%
\\

\bottomrule
\end{tabular}
\end{adjustbox}
\end{table*}



For the top-$k$ recommendation task, we utilize the SELFRec\footnote{\url{https://github.com/Coder-Yu/SELFRec}} library \cite{yu2023self} for implementation.
As for the hyper-parameter settings, 
we set $\alpha_1=\alpha_2=\alpha_3=\frac{1}{3}$ for all experiments. 
$\mathcal{C}$ is set to 0.92 in this paper.
$\mathcal{C}_1$, $\mathcal{C}_2$, and $\mathcal{C}_3$ are set to 0.05, 0.5, and 0.025 respectively. 
We repeat the experiment five times and calculate the average.




\subsection{Main Results (RQ1)}
The experiment results for direct recommendation and sequential recommendation are shown in Table \ref{tab:t2} and Table \ref{tab:t3} respectively.
We have the following key observations:

\begin{itemize}[leftmargin=*]
\item
In the context of MF and LightGCN, pairwise and listwise ranking methods surpass the baseline model. However, these methods encounter difficulties in yielding favorable outcomes when applied to more advanced models like MixGCF or SGL. In contrast, pointwise ranking consistently outperforms the base models, achieving a marked improvement. This enhancement might be attributed to the LLM proficiency in making more objective judgments, rather than comparing multiple items. Additionally, the relative simplicity of pointwise tasks suggests that LLMs are more adept at handling simpler tasks.
\item
Furthermore, hybrid ranking methods generally outperform pointwise ranking. Despite the significantly lower performance of pairwise and listwise ranking compared to pointwise ranking, integrating them into a hybrid ranking approach can still result in improvements. This is in line with the concept of self-consistency in LLMs; that is, when a model consistently agrees on a particular answer, there is a higher likelihood of its accuracy.

\item
\model~demonstrates a more significant improvement on the Bookcrossing dataset than on the Movielens dataset. This enhancement may be due to the fine-grained ratings in Bookcrossing dataset, which range from 1 to 10, thereby enabling the tuned LLM to make more precise predictions.





\end{itemize}

This observation can be attributed to the fact that the general recommendation models have the capability to mine collaborative information effectively, which makes them more excel at ranking items. As a result, the need for reranking is comparatively lower in these models.

\begin{figure*}[t!]
    \centering
    \includegraphics[width=1\linewidth]{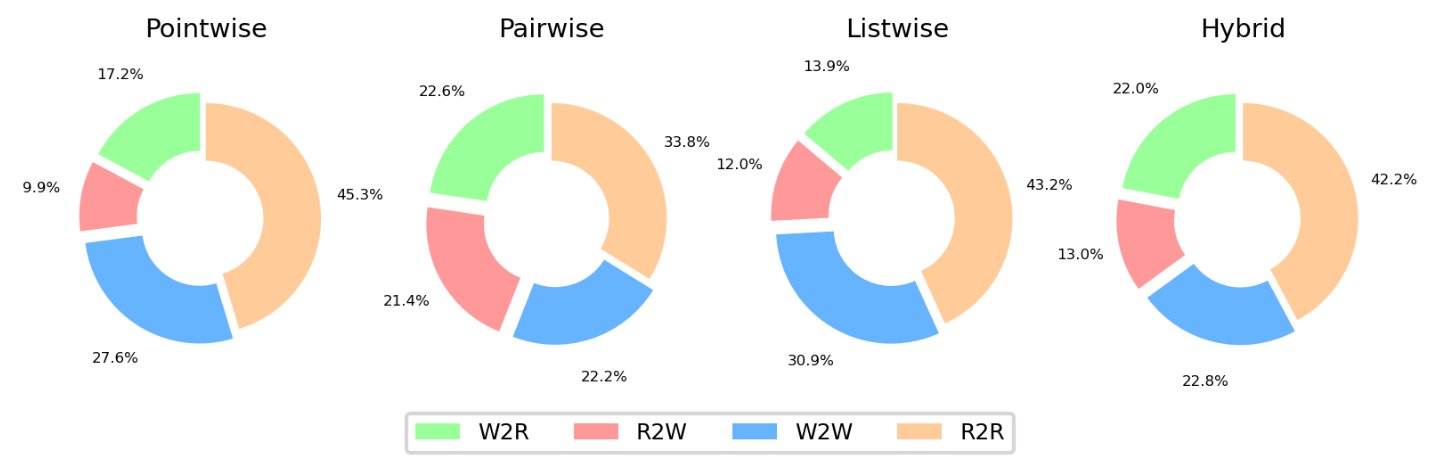}
    \caption{Evaluation of the changes after reranking on ML-1M dataset with backbone model SGL.  W2R: the wrong recommendation is changed to right. R2W: the right recommendation is altered to wrong. W2W: the wrong recommendation remains unchanged. R2R: the correct recommendation remains unchanged.}
    \label{fig:sgl1}
\end{figure*}

\begin{figure*}[t!]
    \centering
    \includegraphics[width=1\linewidth]{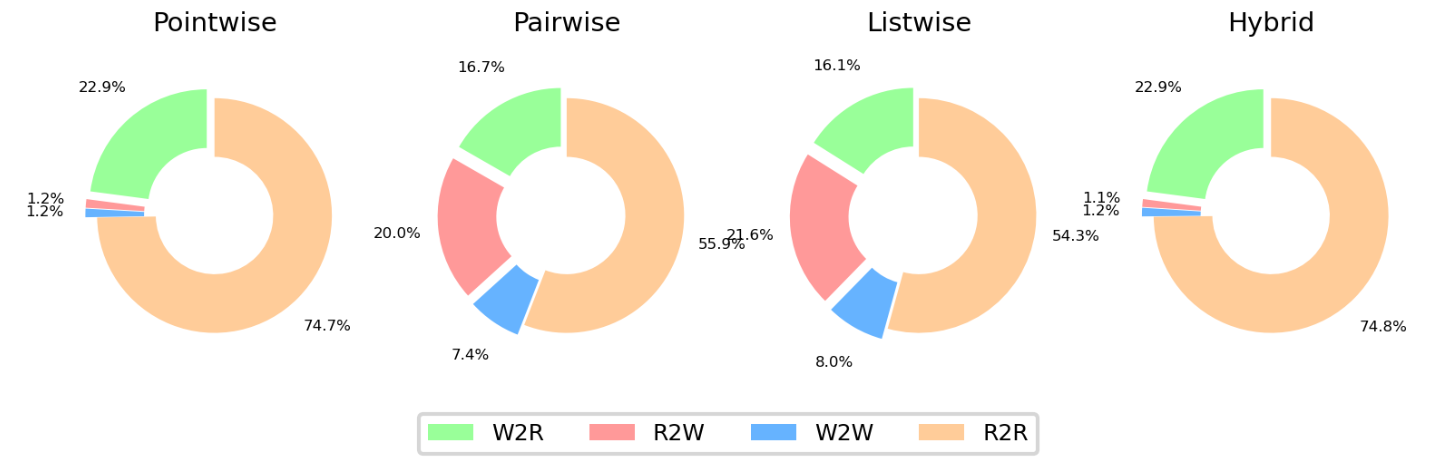}
    \caption{Evaluation of the changes after reranking on Bookcrossing dataset with backbone model SGL.}
    \label{fig:sgl2}
\end{figure*}

\subsection{Model Analysis (RQ2)}
In this section, we analyze the model interns of the efficiency and the reranking.

\subsubsection{Model Efficiency}
In our experiment, we observed the \textbf{training and inference time} of LLM largely exceeds that of the conventional recommendation model.
Specifically, training the LLaMA-2 7B model with around 56K instructions on 16 A800 GPUs took approximately 4.6 hours. Besides, training the LLaMA-2 13B model under the same conditions required around 5.3 hours. The inference time for each instruction averaged about 17 instructions per second, translating to a requirement of around 0.059 seconds per item for computation by a single A800 GPU. 
This finding highlights the limitations of current LLM-based recommender systems. 

\subsubsection{Rerank Analysis}
It is essential to evaluate the \model performance based on the changes it induces during the reranking stage. To do this, we first select results where the ground truth item is included within the top $k'$ recommendations produced by the retrieval model. For the purpose of this evaluation, we define a recommendation as "correct" if the ground truth item appears in the top $k$ positions; otherwise, we categorize the recommendation as "wrong".

As depicted in Figure \ref{fig:sgl1} and \ref{fig:sgl2}, the model demonstrates significant reranking capabilities for top-k recommendations on the Bookcrossing dataset. Over 15\% of the time, the model successfully alters wrong recommendations to correct ones. However, the performance of different ranking methods can vary significantly depending on the dataset. For instance, on the Bookcrossing dataset, the pairwise ranking method tends to change more recommendations from correct to wrong. These results underscore the importance of careful selection of ranking methods, taking into account the specific characteristics of the dataset, to optimize performance.

\begin{table}[t]
\centering
\caption{ Ablation Study on ML-100K dataset with backbone model MF for pairwise ranking. The best results are highlighted in boldfaces.}
\label{tab:ablation}
\begin{tabular}{lcc c c c ccc}
\toprule

{Variants}  & H@3 $\uparrow$ &  N@3 $\uparrow$& H@5 $\uparrow$ &  N@5 $\uparrow$\\ \midrule

\model &\textbf{0.0533} & \textbf{0.0368} &\textbf{0.0783} & \textbf{0.0471} \\ 
\textit{w/o} Adaptive User Sampling&0.0472 & 0.0347 &0.0759 & 0.0465\\ 
\textit{w/o} Position Shifting &0.0472 & 0.0337 &0.0764 & 0.0456\\
\textit{w/o} Prompt Enhancement &0.0494 & 0.0358 &0.0742 & 0.0459\\

\bottomrule
\end{tabular}
\end{table}

\subsection{Ablation Study (RQ3)}
\label{sec:ab}
{
We study the benefits of each individual component of ReRanker. The results are demonstrated in Table~\ref{tab:ablation}.
The results demonstrate that the complete model outperforms all three model variants. 
}
This outcome underscores the significant contribution of each main component to the enhancement of overall performance. A detailed analysis of each component's specific impact yielded the following insights:

\begin{itemize}[leftmargin=*]

\item \textbf{\textit{w/o} Adaptive User Sampling}: 
This variant substitutes the proposed adaptive user sampling with a uniform sampling approach. The experimental results reveal a notable decline in model performance. This decline underscores the importance of adaptive user sampling in selecting critical, representative, and diverse user samples for training, thereby enhancing model performance.

\item \textbf{\textit{w/o} Position Shifting}: 
The position shifting is excluded in this variant, maintaining other components the same. The observed performance reduction in this variant highlights the significance of position shifting. It mitigates position bias, leading to more consistent and reliable results.

\item \textbf{\textit{w/o} Prompt Enhancement}: 
In this variant, prompt enhancement is removed while retaining other modules. A marked decrease in performance is observed, suggesting that conventional recommender models may provide valuable information for LLM to generate more accurate predictions.

\end{itemize}





\subsection{Hyper-parameter Study (RQ4)}
\subsubsection{Analysis of hyper-parameters $\mathcal{C}_1$, $\mathcal{C}_2$ and $\mathcal{C}_3$}
\begin{figure}[t]
\vspace{-8pt}
    \centering
    \subfloat[\centering Impact of $\mathcal{C}_1$]{{\includegraphics[width=.34\linewidth]{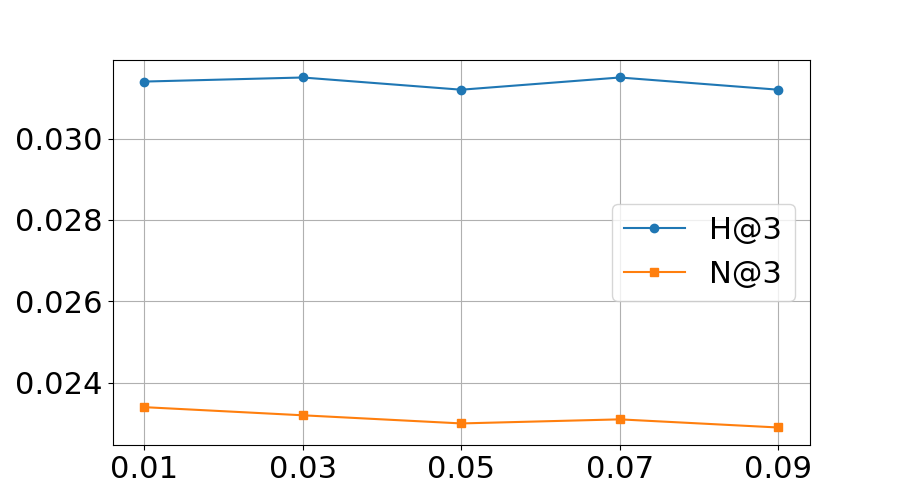} }}%
    \hspace{-11.5pt}
    \subfloat[\centering Impact of $\mathcal{C}_2$]{{\includegraphics[width=.34\linewidth]{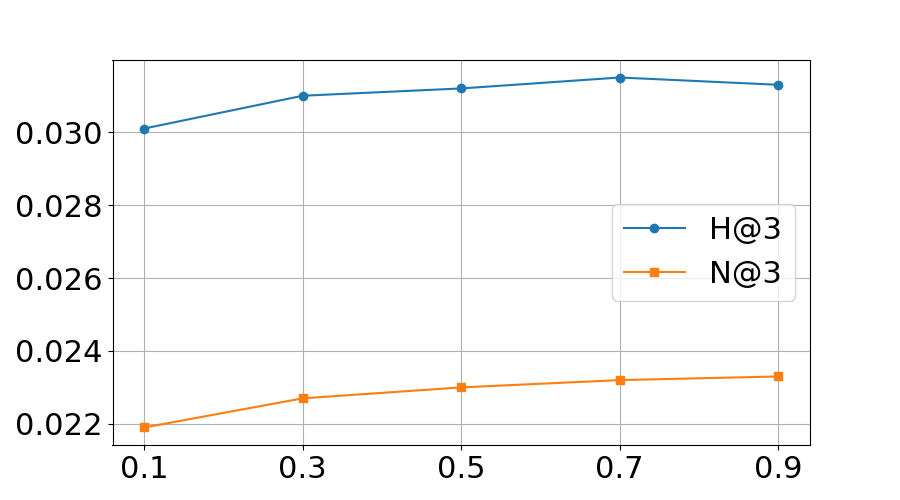} }}%
    \hspace{-11.5pt}
    \subfloat[\centering Impact of $\mathcal{C}_3$]{{\includegraphics[width=.34\linewidth]{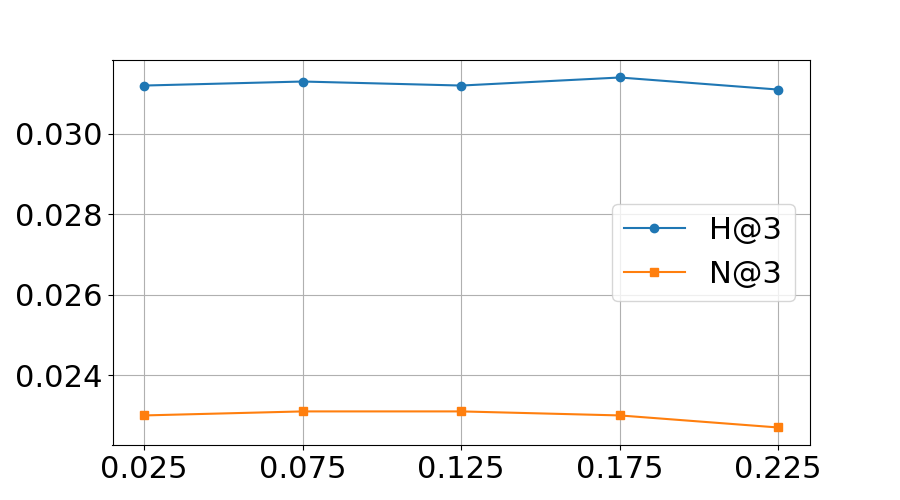} }}%
    \caption{
    Analysis of hyper-parameters $\mathcal{C}_1$, $\mathcal{C}_2$ and $\mathcal{C}_3$ on ML-1M dataset with backbone model MF and hybrid ranking task.}%
    \label{fig:f3}%
\end{figure}

We analyze the influence of hyper-parameters $\mathcal{C}_1$, $\mathcal{C}_2$, and $\mathcal{C}_3$ on the ML-1M dataset, employing MF as the underlying model, as depicted in Figure \ref{fig:f3}. We noted that increases in $\mathcal{C}_1$ and $\mathcal{C}_3$ led to fluctuations and a general decline in performance. This indicates that judicious selection of $\mathcal{C}_1$ and $\mathcal{C}_3$ is crucial for optimizing model performance, particularly since both pairwise and listwise ranking methods underperform compared to pointwise ranking, rendering high values of $\mathcal{C}_1$ and $\mathcal{C}_3$ suboptimal. On the other hand, a gradual improvement in performance was observed with the increment of $\mathcal{C}_2$. These findings underscore the significance of appropriate hyper-parameter selection in achieving optimal model performance.

\begin{figure*}%
\vspace{-7pt}
    \centering
    \subfloat[\centering Pointwise Ranking]{{\includegraphics[width=.24\linewidth]{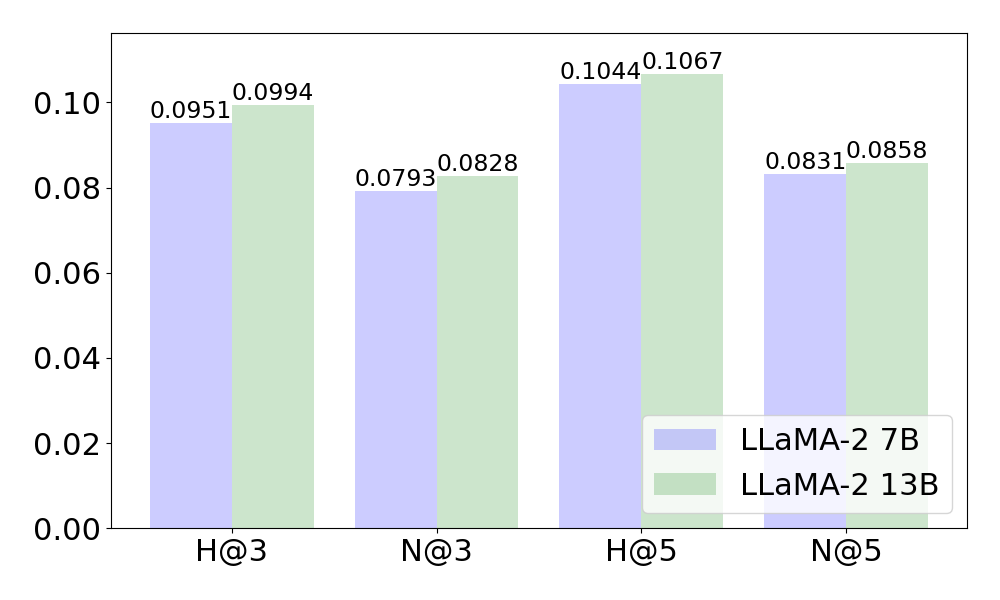} }}%
    \subfloat[\centering Pairwise Ranking]{{\includegraphics[width=.24\linewidth]{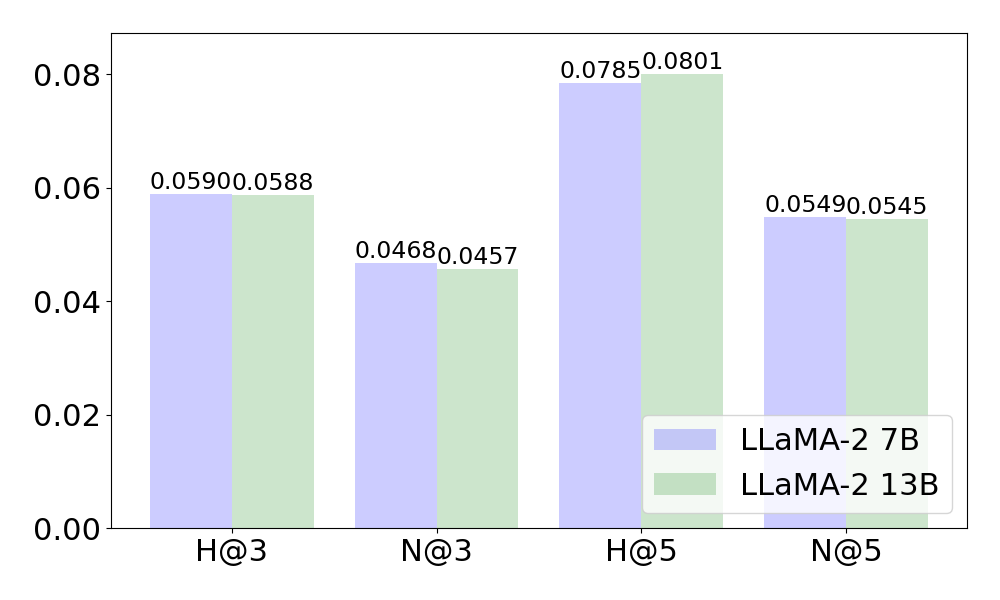} }}%
    \subfloat[\centering Listwise Ranking]{{\includegraphics[width=.24\linewidth]{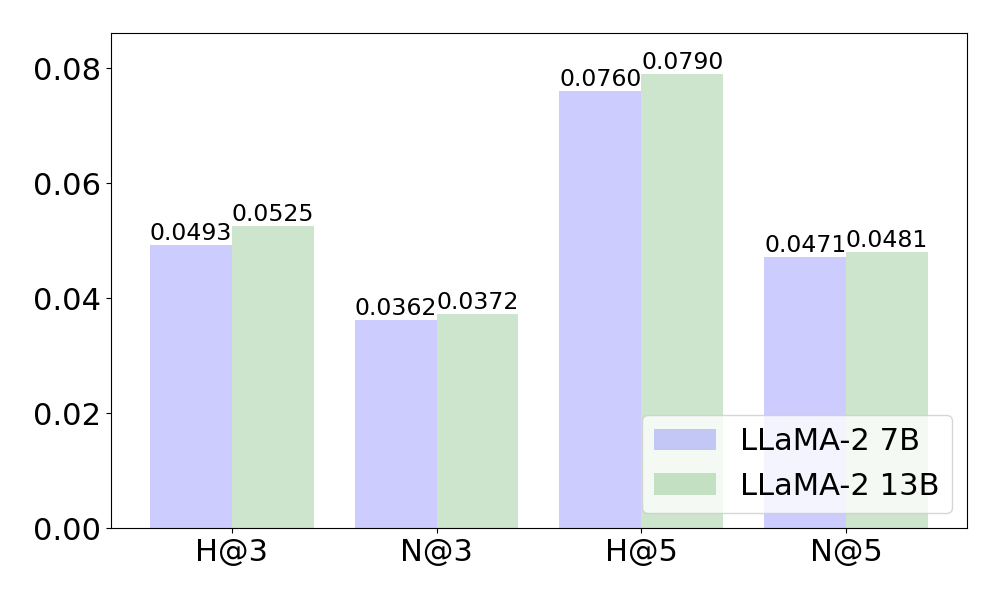} }}%
    \subfloat[\centering Hybrid Ranking]{{\includegraphics[width=.24\linewidth]{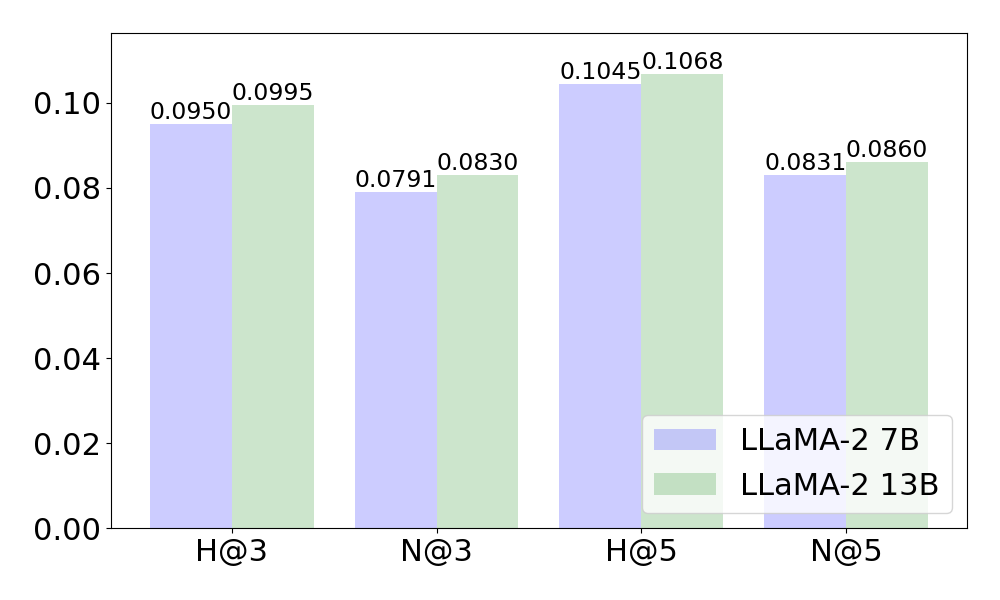} }}%
    \caption{Performance comparison for LLaMA-2 (7B) and LLaMA-2 (13B) model with respect to different ranking tasks on the Bookcrossing dataset with backbone model SGL.}%
    \label{fig:f4}%
\end{figure*}

\subsubsection{Analysis of Model Scaling.}

\begin{table}[t]
\centering
\caption{ 
Performance comparison w.r.t different numbers of instructions for training \model~on the ML-100K dataset using the MF backbone model for pairwise ranking}
\label{tab:data}
\begin{tabular}{lcc c c c ccc}
\toprule

{\# of Instructions}  & H@3 $\uparrow$ &  N@3 $\uparrow$& H@5 $\uparrow$ &  N@5 $\uparrow$\\ \midrule

56K &\textbf{0.0533} & \textbf{0.0368} &\textbf{0.0783} & \textbf{0.0471} \\ 
28K&0.0481 & 0.0348 &0.0757 & 0.0462\\ 
5.6K &0.0475 & 0.0353 &0.0723 & 0.0454\\

\bottomrule
\end{tabular}
\end{table}

We {further} instruction-tuned the LLaMA-2 (13B) model.\footnote{
Training the LLaMA-2 (70B) model with the same experimental settings was impractical due to resource constraints, consistently resulting in Out-Of-Memory (OOM) errors.}
We conducted a comparative analysis between the 7B and 13B versions of the instruction-tuned models. The performance differences between LLaMA-2 7B and LLaMA-2 13B were specifically assessed across various ranking tasks within the Bookcrossing dataset, as illustrated in Figure~\ref{fig:f4}. Our observations revealed that the LLaMA-2 (13B) model generally outperformed the 7B model. This superiority can be attributed to the enhanced capabilities of the larger model, which result in better language comprehension and reasoning ability, ultimately leading to improved ranking outcomes. In addition,
it is noteworthy that the improvements in pointwise ranking and listwise ranking were more pronounced compared to pairwise ranking. This suggests that LLMs still face challenges in certain ranking tasks. Furthermore, the hybrid ranking approach demonstrated significant progress across all evaluation metrics. This underscores the effectiveness of integrating multiple ranking tasks, highlighting the strengths of the proposed hybrid ranking method.

\subsubsection{Analysis of Data Scaling.}
The training of the LLM was conducted with varying quantities of instructions in the instruction-tuning dataset to evaluate the effect of data size. Specifically, the version with 5.6K instructions was trained over 600 steps, while the version with 28K instructions underwent 3000 steps of training, proportional to our original configuration. The experiment result is detailed in Table~\ref{tab:data}. An observable trend is that an increase in the number of instructions correlates with enhanced model performance. This underscores the significance of incorporating a larger and more diverse dataset for instruction tuning LLMs to achieve improved performance.






\begin{figure}[]
    \centering 
\includegraphics[width=0.97\linewidth]{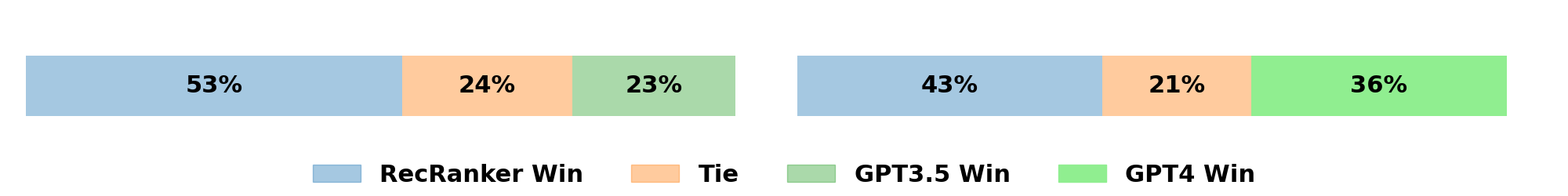}
    \caption{Comparison between our instruction-tuned model with the GPT-3.5-turbo and GPT-4 model.
    }
\label{fig:comparegpt}
\vspace{-0.05in}
\end{figure}

\subsection{Comparison with the GPT Model (RQ5)}
We compare our instruction-tuned LLM with the GPT model, specifically, the GPT-3.5-turbo\footnote{\url{https://platform.openai.com/docs/models/gpt-3-5}} and GPT-4\footnote{\url{https://platform.openai.com/docs/models/gpt-4}} model. 
We employed a sample of 100 listwise ranking task instances from the Bookcrossing dataset, using the CLSRec model as the backbone for evaluating the GPT model.
This experiment setting aligns with the findings of \cite{dai2023uncovering}, which highlight the optimal cost-performance equilibrium achieved when GPT-3.5 is applied to the listwise ranking task.
The experimental result is shown in Figure \ref{fig:comparegpt}.
We could observe that our instruction-tuned \model~with hybrid ranking notably outperforms the GPT-3.5 model. 
Moreover, the GPT-4 model perform much better than GPT-3.5 model, while still under-perform \model.
This impressive result emphasizes the crucial role of instruction tuning in aligning general-purpose LLMs specifically for recommendation tasks.


\subsection{Further Discussion}

In this section, we discuss the potential application and further improvement of LLM-based recommendations.
We could observe the training and inference duration significantly exceeds that of conventional recommendation models, highlighting the limitations of current LLM-based recommender systems. 
The substantial demand for computational resources also represents a significant challenge.
Consequently, employing instruction LLMs for large-scale industrial recommender systems, such as those with millions of users, is presently impractical. However, future advancements in accelerated and parallel computing algorithms for language model inference could potentially reduce inference times and computation resources. This improvement might make the integration of LLMs into large-scale recommender systems feasible, especially by leveraging many GPUs for parallel computation \cite{shoeybi2019megatron}.
Moreover, another possible solution is model distillation \cite{liang2020mixkd}.
By distilling a smaller model from the larger LLM, we can potentially retain the powerful predictive capabilities of the LLM while significantly reducing the computational resources required for training and inference.
Furthermore, the distilled model is not only computationally cheaper to run, but it's also faster at generating recommendations, making it more suitable for real-time applications.
In addition to model distillation, other techniques such as quantization \cite{bai2022towards} and pruning \cite{ma2023llm} can also be used to reduce the size and complexity of LLMs without significantly compromising their performance. These methods could be combined together to further optimize the performance and efficiency of LLM-based recommender systems.

\section{Conclusion}

In this paper, we introduce \textbf{\model}, a novel framework for employing instruction tuning large {la}nguage {m}odel {a}s the \textbf{Ranker} in top-\textit{k} \textbf{Rec}ommendations. Initially, we propose an adaptive user sampling module for obtaining high-quality, representative, and diverse data.
Afterward, we construct an instruction-tuning dataset that encompasses three distinct ranking tasks: pointwise, pairwise, and listwise rankings. We further improve the prompt by 
adopting position shifting strategy to mitigate position bias, as well as integrating auxiliary information from conventional recommendation models for prompt enhancement.
Moreover, we introduce a hybrid ranking method that combines these diverse ranking tasks to improve overall model performance.
Extensive empirical studies on three real-world datasets across diverse rankings tasks validate the effectiveness of our proposed framework. 


\bibliographystyle{ACM-Reference-Format}
\bibliography{ref}

\end{document}